\documentclass{article}
\usepackage{graphicx}
\usepackage{amsmath}

\begin{document}

\title{\textbf{The way back: from charge conservation to Maxwell equations}}
\author{F. De\ Zela \\
Departamento de Ciencias, Secci\'{o}n F\'{i}sica \\
Pontificia Universidad Cat\'{o}lica del Per\'{u}, Ap.1761, Lima,
Per\'{u}.\\
 fdezela@fisica.pucp.edu.pe} \maketitle

\begin{abstract}
The main purpose of this article is to disseminate among a wide audience of
physicists a known result, which is available since a couple of years to the
\emph{cognoscenti} of differential forms on manifolds; namely, that charge
conservation implies the inhomogeneous Maxwell equations. This is the
reciprocal statement of one which is very well known among physicists:
charge conservation, written in the form of a continuity equation, follows
as a consequence of Maxwell equations. We discuss the conditions under which
charge conservation implies Maxwell equations. The key role played by the
constitutive equations is hereby stressed. The discussion is based on
Helmholtz theorem, according to which a vector field is determined by its
divergence and its curl. Green's functions are also shown to play a
fundamental role. We present all results in three-vector, as well as in
tensorial notation. We employ only those mathematical tools most physicists
are familiar with.
\end{abstract}

\newpage

\section{Introduction}

Maxwell equations are frequently introduced - using gaussian units - in the
following form \cite{jackson}:

\begin{eqnarray}
\mathbf{\nabla }\cdot \mathbf{D} &=&4\pi \rho  \label{1} \\
\mathbf{\nabla }\times \mathbf{H-}\frac{1}{c}\frac{\partial \mathbf{D}}{%
\partial t} &=&\frac{4\pi }{c}\mathbf{j}  \label{2} \\
\mathbf{\nabla }\cdot \mathbf{B} &=&0  \label{3} \\
\mathbf{\nabla }\times \mathbf{E}+\frac{1}{c}\frac{\partial \mathbf{B}}{%
\partial t} &=&\mathbf{0}.  \label{4}
\end{eqnarray}

Equations (\ref{1}) and (\ref{2}) are called inhomogeneous - or Maxwell
equations with sources -, while (\ref{3}) and (\ref{4}) are called
homogeneous, or source-free equations. The four equations constitute a
closed system because the couples $(\mathbf{D},\mathbf{H})$ and $(\mathbf{E},%
\mathbf{B})$ are related to each other through the so-called ``constitutive
equations''. It is however not unusual to stress the fact that the equations
with sources are, to some extent, conceptually different from the
source-free equations. Indeed, Eqs.(\ref{3}) and (\ref{4}) can be understood
as expressing a purely mathematical statement. To see this we start by
considering a scalar field $\varphi (t,\mathbf{x})$ and a vector field $%
\mathbf{A}(t,\mathbf{x})$, which are continuously differentiable but
otherwise totally arbitrary. Then we construct the fields $\mathbf{B}\equiv
\mathbf{\nabla }\times \mathbf{A}$ and $\mathbf{E}\equiv -\mathbf{\nabla }%
\varphi -\partial _{t}\mathbf{A}/c$. Eq.(\ref{3}) is then identically
satisfied because the divergence of a curl vanishes. If we now take the curl
of $\mathbf{E}$ and use the fact that the curl of a gradient vanishes, we
see that Eq.(\ref{4}) also holds true identically. We conclude that Eqs.(\ref
{3}) and (\ref{4}) are satisfied by \emph{arbitrary} fields, as long as
these fields are constructed as above, starting from the given fields $%
\varphi $ and $\mathbf{A}$. These two equations are therefore not
characteristic of the electromagnetic field. They can be understood as a
mathematical statement telling us that there are fields, $\varphi $ and $%
\mathbf{A}$, out of which we can construct $\mathbf{E}$ and $\mathbf{B}$.
The electromagnetic nature of these fields depends on the fact that they
have to satisfy also equations (\ref{1}) and (\ref{2}), as long as $\mathbf{D%
}=\mathbf{D}(\mathbf{E},\mathbf{B})$ and $\mathbf{H}=\mathbf{H}(\mathbf{E},%
\mathbf{B})$. Equations (\ref{1}) and (\ref{2}) are the ones possessing a
truly physical content. They are the ones which contain the sources that
produce the field. It is the particular way in which these sources are
related to the fields, what makes up the physical content of these equations.

Let us now turn to Maxwell equations as they are often written in tensorial
form:

\begin{eqnarray}
\partial _{\mu }F^{\mu \nu } &=&\frac{4\pi }{c}j^{\nu },  \label{5} \\
\partial _{\mu }{}F_{\nu \lambda }+\partial _{\nu }{}F_{\lambda \mu
}+\partial _{\lambda }{}F_{\mu \nu } &=&0.  \label{6}
\end{eqnarray}

Here again, starting from an \emph{arbitrary} four-vector $A_{\mu }(x)$ we
may define an antisymmetric tensor $F_{\mu \nu }\equiv \partial _{\mu
}{}A_{\nu }-\partial _{\nu }{}A_{\mu }$. It is easy to see that this tensor
identically satisfies the homogeneous equation (\ref{6}), which is a Bianchi
type identity. As before, if our $A_{\mu }$ has to describe an
electromagnetic field, then it has to satisfy the inhomogeneous equation $%
\partial _{\mu }\partial ^{\mu }A^{\nu }-\partial ^{\nu }\partial _{\mu
}A^{\mu }=4\pi j^{\nu }/c$, which is another form of Eq.(\ref{5}).

Summarizing, we can say that the homogeneous Maxwell equations can be
considered as entailing a mathematical statement about the nature of the
fields $\mathbf{E}$ and $\mathbf{B}$, or -correspondingly - about the tensor
$F_{\mu \nu }$. The inhomogeneous Maxwell equations in turn are the ones
possessing physical content. We must postulate that the electromagnetic
field has to satisfy them. Now, all these things are very well known. What
seems to be not so very well known is the fact that the inhomogeneous
equations by themselves are also \emph{not} characteristic of the
electromagnetic field. Indeed, suppose we are given a scalar function $\rho
(t,\mathbf{r})$ and a vector function $\mathbf{j}(t,\mathbf{r})$, both of
which go to zero sufficiently rapidly as $r\rightarrow \infty $, and being
such that they satisfy the equation

\begin{equation}
\partial _{t}\rho +\mathbf{\nabla }\cdot \mathbf{j}=0.  \label{7}
\end{equation}

Then there exist vector fields $\mathbf{D}(t,\mathbf{r})$ and $\mathbf{H}(t,%
\mathbf{r})$ satisfying the inhomogeneous Maxwell equations

\begin{eqnarray}
\mathbf{\nabla }\cdot \mathbf{D} &=&\rho  \label{8} \\
\mathbf{\nabla }\times \mathbf{H} &=&\mathbf{j}+\partial _{t}\mathbf{D.}
\label{9}
\end{eqnarray}

The existence of $\mathbf{D}$ and $\mathbf{H}$ can be proved by explicit
construction. Such a construction rests on Helmholtz theorem \cite
{hauser,kobe1,kobe2,baierlein,arfken}, which is discussed below. For now, it
suffices to say that - loosely speaking - ``a vector field is determined by
its divergence and its curl''. Thus, according to Helmholtz theorem, Eq.(\ref
{8}) can be solved for $\mathbf{D}$ (though the solution is not unique) when
$\rho $ is given. From equations (\ref{8}) and (\ref{7}) we see that $%
\mathbf{\nabla }\cdot \left( \partial _{t}\mathbf{D}+\mathbf{j}\right) =0$.
Applying Helmholtz theorem again we can show that there is a field $\mathbf{H%
}$ whose curl is $\partial _{t}\mathbf{D}+\mathbf{j}$. This is equation (\ref
{9}). Note that we have written the inhomogeneous Maxwell equations in MKS
units, which are the convenient units for what follows.

We see then that the continuity equation (\ref{7}) entails the inhomogeneous
Maxwell equations. The continuity equation expresses the conservation of
something. This something must not necessarily be electric charge. It could
be mass as well, or any other quantity - like probability, for instance. We
are thus led to conclude that the inhomogeneous Maxwell equations are also
not characteristic of electromagnetism. They hold true whenever something is
conserved. Putting things this way we bring to the fore the fundamental role
played by the constitutive equations, $\mathbf{D}=\mathbf{D}(\mathbf{E},%
\mathbf{B})$ and $\mathbf{H}=\mathbf{H}(\mathbf{E},\mathbf{B})$, whatever
their precise form might be. They constitute the link between the
homogeneous and the inhomogeneous Maxwell equations. It is this link what
turns the four equations into a closed system. Neither the inhomogeneous nor
the homogeneous equations by themselves are characteristic of
electromagnetism. They must be linked to one another in order to conform a
closed system of equations that is characteristic of electromagnetic
phenomena.

In the following section we discuss Helmholtz theorem. Although this theorem
can be found in several textbooks and articles, for our purposes it is
useful to present it in a form which brings to the fore its connection with
Green's functions.

\section{Helmholtz theorem}

Here we discuss Helmholtz theorem by following an approach which is slightly
different from the one presented in several textbooks. Helmholtz theorem
states that a vector field $\mathbf{v}$ is completely determined by giving
its divergence and its curl, together with its normal component, $\widehat{%
\mathbf{n}}\cdot \mathbf{v}$, at the boundary of the domain where such a
vector field is to be determined. For physical applications it is natural to
take as ``boundary'' an infinitely distant surface and $\mathbf{v}$
vanishing there. Helmholtz theorem then says that we can write $\mathbf{v}$
in terms of two potentials, $U$ and $\mathbf{C}$, in the form

\begin{equation}
\mathbf{v(x)}=\mathbf{\nabla }U\mathbf{(x)}+\mathbf{\nabla }\times \mathbf{%
C(x)},  \label{10}
\end{equation}
where $U$ and $\mathbf{C}$ can be expressed in terms of the divergence and
the curl of $\mathbf{v}(\mathbf{r})$, respectively. Now, put in this form,
Helmholtz's theorem might appear as a result that is rather awkward to
prove. Let us thus try to lay bare what motivates it. To this end, consider
the following two vector identities, in which the Laplacian $\mathbf{\nabla }%
^{2}$ appears:

\begin{eqnarray}
\mathbf{\nabla \cdot }\left( \mathbf{\nabla }U\right) &=&\mathbf{\nabla }%
^{2}U  \label{11} \\
\mathbf{\nabla }\times \left( \mathbf{\nabla }\times \mathbf{C}\right) &=&%
\mathbf{\nabla }\left( \mathbf{\nabla \cdot C}\right) -\mathbf{\nabla }^{2}%
\mathbf{C}.  \label{12}
\end{eqnarray}

Add to these relations the equation satisfied by a Green function $G(\mathbf{%
x},\mathbf{y})$, on which we impose the condition that it vanishes at
infinity:

\begin{eqnarray}
\mathbf{\nabla }^{2}G(\mathbf{x},\mathbf{y}) &=&\delta ^{3}(\mathbf{x}-%
\mathbf{y}),  \label{13} \\
G(\mathbf{x},\mathbf{y}) &=&-\frac{1}{4\pi \left| \mathbf{x}-\mathbf{y}%
\right| }.  \label{14}
\end{eqnarray}

By means of $G(\mathbf{x},\mathbf{y})$ we can introduce $U$ and $\mathbf{C}$
as ``potentials'' associated with two given ``densities'', $\rho $ and $%
\mathbf{j}$, through

\begin{eqnarray}
U(\mathbf{x}) &=&-\frac{1}{4\pi }\int \frac{\rho (\mathbf{y})}{\left|
\mathbf{x}-\mathbf{y}\right| }d^{3}y,  \label{15} \\
\mathbf{C}(\mathbf{x}) &=&\frac{1}{4\pi }\int \frac{\mathbf{j}(\mathbf{y})}{%
\left| \mathbf{x}-\mathbf{y}\right| }d^{3}y.  \label{16}
\end{eqnarray}

We assume that $\rho $ and $\mathbf{j}$ vanish at infinity. The potentials
then satisfy

\begin{eqnarray}
\mathbf{\nabla }^{2}U(\mathbf{x}) &=&\rho (\mathbf{x}),\text{ }  \label{17}
\\
\mathbf{\nabla }^{2}\mathbf{C}(\mathbf{x}) &=&-\mathbf{j}(\mathbf{x}).
\label{17a} \\
\text{ }\mathbf{\nabla \cdot j(x)} &=&0\Rightarrow \mathbf{\nabla \cdot C(x)}%
=0.  \label{18}
\end{eqnarray}

The validity of Eqs.(\ref{17},\ref{17a}) follows directly from the
definitions given by Eqs.(\ref{15},\ref{16}), together with Eqs.(\ref{13},%
\ref{14}). In order to see that $\mathbf{\nabla \cdot j(x)}=0$ implies that $%
\mathbf{C(x)}$ is divergenless we need a little more ellaborated calculation:

\begin{eqnarray}
\mathbf{\nabla \cdot C(x)} &=&\frac{1}{4\pi }\int \mathbf{\nabla }_{\mathbf{x%
}}\mathbf{\cdot }\left( \frac{\mathbf{j}(\mathbf{y})}{\left| \mathbf{x}-%
\mathbf{y}\right| }\right) d^{3}y=\frac{1}{4\pi }\int \mathbf{j}(\mathbf{y}%
)\cdot \mathbf{\nabla }_{\mathbf{x}}\left( \frac{1}{\left| \mathbf{x}-%
\mathbf{y}\right| }\right) d^{3}y  \label{19} \\
&=&-\frac{1}{4\pi }\int \mathbf{j}(\mathbf{y})\cdot \mathbf{\nabla }_{%
\mathbf{y}}\left( \frac{1}{\left| \mathbf{x}-\mathbf{y}\right| }\right)
d^{3}y  \label{20} \\
&=&-\frac{1}{4\pi }\int \sum_{i=1}^{3}j_{i}(\mathbf{y})\frac{\partial }{%
\partial y^{i}}\left( \frac{1}{\left| \mathbf{x}-\mathbf{y}\right| }\right)
d^{3}y \\
&=&\frac{-1}{4\pi }\left[ \int \frac{\mathbf{j}(\mathbf{y})\cdot \mathbf{n}(%
\mathbf{y})}{\left| \mathbf{x}-\mathbf{y}\right| }dS-\int \sum_{i=1}^{3}%
\frac{1}{\left| \mathbf{x}-\mathbf{y}\right| }\frac{\partial j_{i}(\mathbf{y}%
)}{\partial y^{i}}d^{3}y\right] =0.
\end{eqnarray}

In the last step - which resulted from an integration by parts - the volume
integral was replaced by a surface integral using the divergence - or Stokes
- theorem. Such a surface integral vanishes when the volume of integration
goes to infinity, because $\mathbf{j}$ has been assumed to vanish at
infinity. The second term vanishes because of the requirement $\mathbf{%
\nabla \cdot j}=0$.

From Eqs.(\ref{11}) and (\ref{12}) together with (\ref{17},\ref{17a},\ref{18}%
), we see that

\begin{eqnarray}
\mathbf{\nabla \cdot }\left( \mathbf{\nabla }U\right) &=&\rho (\mathbf{x}),
\\
\mathbf{\nabla }\times \left( \mathbf{\nabla }\times \mathbf{C}\right) &=&%
\mathbf{j}(\mathbf{x}).
\end{eqnarray}

This suggests us to define a field $\mathbf{v}=\mathbf{\nabla }U+\mathbf{%
\nabla }\times \mathbf{C}$. Such a field satisfies

\begin{equation}
\mathbf{\nabla \cdot v}=\rho (\mathbf{x}),\text{ }\mathbf{\nabla }\times
\mathbf{v}=\mathbf{j}(\mathbf{x}).
\end{equation}

This way we arrive naturally at the following statement: if we are given the
divergence $\rho (\mathbf{x})$ and the curl $\mathbf{j}(\mathbf{x})$ of a
vector field $\mathbf{v}(\mathbf{x})$ which vanishes at infinity, then we
can write this field as $\mathbf{v}=\mathbf{\nabla }U+\mathbf{\nabla }\times
\mathbf{C}$, where $U$ and $\mathbf{C}$ are given in terms of $\rho $ and $%
\mathbf{j}$ by Eqs.(\ref{15}) and (\ref{16}). In order to see that $\rho $
and $\mathbf{j}$ uniquely determine $\mathbf{v}$, it suffices to show that
when both the divergence and the curl of a field vanish, then the field
itself vanishes identically. This follows from what we have done so far.
Indeed, we have shown that the following equation holds true identically:

\begin{equation}
\mathbf{v}(\mathbf{x})=-\mathbf{\nabla }_{\mathbf{x}}\left( \int G(\mathbf{x}%
,\mathbf{y})\mathbf{\nabla }\cdot \mathbf{v}(\mathbf{y})d^{3}y\right) +%
\mathbf{\nabla }_{\mathbf{x}}\times \left( \int G(\mathbf{x},\mathbf{y})%
\mathbf{\nabla }\times \mathbf{v}(\mathbf{y})d^{3}y\right) ,  \label{25}
\end{equation}
with the Green's function $G(\mathbf{x},\mathbf{y})$ satisfying Eqs.(\ref{13}%
,\ref{14}). Hence, if $\mathbf{\nabla }\cdot \mathbf{v}=0$ and $\mathbf{%
\nabla }\times \mathbf{v}=\mathbf{0}$, then $\mathbf{v}=\mathbf{0}$. We
conclude that given two fields, $\mathbf{v}_{1}$ and $\mathbf{v}_{2}$,
having the same divergence and curl, they must in fact be the same field.
This, because their difference $\mathbf{v}=\mathbf{v}_{1}-\mathbf{v}_{2}$
vanishes identically, as a consequence of $\mathbf{\nabla }\cdot \mathbf{v}%
=0 $ and $\mathbf{\nabla }\times \mathbf{v}=\mathbf{0}$.

Finally, let us first note that Eq.(\ref{25}) holds for Green's functions
other than the one defined in Eq.(\ref{14}). Indeed, the only property we
need to assume about the Green function $G(\mathbf{x},\mathbf{y})$ is that
it be of the form $G(\mathbf{x}-\mathbf{y})$. This is true anyway, whenever $%
G(\mathbf{x},\mathbf{y})$ fulfills Eq.(\ref{13}). As to the field $\mathbf{v}
$, it has been assumed to vanish at infinity. In fact, it suffices to assume
that it vanishes faster than $1/r$ for large $r$. Note also that if we
prescribe only the divergence $\mathbf{\nabla \cdot v}=\rho (\mathbf{x})$ of
a field, then what we can deduce from this sole condition is that

\begin{equation}
\mathbf{v}(\mathbf{x})=-\mathbf{\nabla }_{\mathbf{x}}\left( \int G(\mathbf{x}%
,\mathbf{y})\mathbf{\nabla }\cdot \mathbf{v}(\mathbf{y})d^{3}y\right) +%
\mathbf{\nabla }_{\mathbf{x}}\times \mathbf{Z(x)},  \label{26}
\end{equation}
with $\mathbf{Z(x)}$ arbitrary. If we instead prescribe only the curl $%
\mathbf{\nabla }\times \mathbf{v}=\mathbf{j}(\mathbf{x})$ of a field, then
we have

\begin{equation}
\mathbf{v}(\mathbf{x})=\mathbf{\nabla }_{\mathbf{x}}\times \left( \int G(%
\mathbf{x},\mathbf{y})\mathbf{\nabla }\times \mathbf{v}(\mathbf{y}%
)d^{3}y\right) +\mathbf{\nabla }_{\mathbf{x}}V(\mathbf{x}),  \label{27}
\end{equation}
with $V\mathbf{(x)}$ arbitrary.

\section{Maxwell equations and Helmoltz theorem}

We have discussed Helmholtz theorem in the framework of $R^{3}$. That is,
the vector fields we have considered are of the form $\mathbf{v}(\mathbf{x})$%
. However, all the results we have obtained so far remain valid if we assume
these fields to depend on a set of additional parameters. They can be
assumed to have been there all the way, but without having been shown
explicitly. Let us denote one of these parameters as $t$. For the moment, we
do not assign to it any physical meaning. Of course, the notation
anticipates that it will be identified in due course with the time variable.

Let us start by assuming that we are given the divergence $\rho $ of a
field, which is a function not only of position but of the parameter $t$ as
well, which we now make explicit, i.e., $\rho =\rho (t,\mathbf{r})$. Let our
boundary condition be such that $\rho $ vanishes at spatial infinity.
Helmholtz theorem states that there is a field, call it $\mathbf{D}$,
satisfying

\begin{equation}
\mathbf{\nabla }\cdot \mathbf{D}(t,\mathbf{r})=\rho (t,\mathbf{r}).
\label{28}
\end{equation}
\

As we have seen, the field $\mathbf{D}(t,\mathbf{r})$ is explicitly given by

\begin{equation}
\mathbf{D}(t,\mathbf{r})=-\mathbf{\nabla }_{\mathbf{r}}\int \frac{\rho (t,%
\mathbf{r}_{1})}{4\pi \left| \mathbf{r}-\mathbf{r}_{1}\right| }d^{3}r_{1}+%
\mathbf{\nabla }_{\mathbf{r}}\times \mathbf{Z}(t,\mathbf{r}),  \label{29}
\end{equation}
with $\mathbf{Z}$ an arbitrary field that we are free to put equal to zero,
if we want. We stress that $t$ plays, in all of this, only the role of a
parameter that can be appended to the fields, \emph{without having any
dynamical meaning}. The field $\mathbf{D}(t,\mathbf{r})$ is required to
satisfy only one condition we have put upon it, i.e., $\mathbf{\nabla }\cdot
\mathbf{D}(t,\mathbf{r})=\rho (t,\mathbf{r}).$ The curl of $\mathbf{D}$ has
been left unspecified, or else set equal to zero.

Consider now a field $\mathbf{j}(t,\mathbf{r})$ depending on the same
parameter $t$ as $\rho $ does. Assume next that $\rho (t,\mathbf{r})$ and $%
\mathbf{j}(t,\mathbf{r})$ satisfy a continuity equation:

\begin{equation}
\partial _{t}\rho +\mathbf{\nabla }\cdot \mathbf{j}=0.  \label{30}
\end{equation}

By using Eq.(\ref{28}) the continuity equation can be written as

\begin{equation}
\mathbf{\nabla }\cdot \left( \partial _{t}\mathbf{D}+\mathbf{j}\right) =0.
\label{31}
\end{equation}

The divergenless vector $\partial _t\mathbf{D}+\mathbf{j}$ can thus be taken
as being the curl of a field $\mathbf{H}(t,\mathbf{r})$. Indeed, according
to what we have seen before, the equation $\mathbf{\nabla }\times \mathbf{H}=%
\mathbf{j}+\partial _t\mathbf{D}$ can be solved as

\begin{equation}
\mathbf{H}(t,\mathbf{r})=\mathbf{\nabla }_{\mathbf{r}}\times \int \frac{%
\mathbf{j(}t,\mathbf{r}_{1}\mathbf{)}+\partial _{t}\mathbf{D(}t,\mathbf{r}%
_{1}\mathbf{)}}{4\pi \left| \mathbf{r}-\mathbf{r}_{1}\right| }d^{3}r_{1}+%
\mathbf{\nabla }_{\mathbf{r}}V(t,\mathbf{r}).  \label{32}
\end{equation}

As long as we do not specify $\mathbf{\nabla }\cdot \mathbf{H}$ the function
$V$ remains undetermined. In any case, the Maxwell equations $\mathbf{\nabla
}\cdot \mathbf{D}=\rho $ and $\mathbf{\nabla }\times \mathbf{H}-\partial _{t}%
\mathbf{D}=\mathbf{j}$ hold true as a consequence of the continuity equation
and Helmholtz theorem. However, these equations are not enough to determine
the dynamics of the fields $\mathbf{D}$ and $\mathbf{H}$, even though we may
ascribe to $t$ the meaning of time. This must be so because - to begin with
- the continuity equation alone does not entail enough information about the
dynamics of $\rho $ and $\mathbf{j}$. But even in case we were provided with
the complete dynamics of $\rho $ and $\mathbf{j}$, from a physical point of
view it is clear that some assumptions must be made concerning the
properties of the medium (e.g., ``space-time'') in order to fix the dynamics
of the electromagnetic fields that will eventually propagate in such a
medium.

Nonetheless, let us pursue a little bit further the mathematical approach
suggested by Helmholtz theorem. The potentials $U$ and $\mathbf{C}$ in terms
of which we defined the field $\mathbf{v}(\mathbf{x})$ read here

\begin{eqnarray}
\varphi (t,\mathbf{r}) &=&\int \frac{\rho \mathbf{(}t,\mathbf{r}_{1}\mathbf{)%
}}{4\pi \left| \mathbf{r}-\mathbf{r}_{1}\right| }d^{3}r_{1},  \label{33} \\
\mathbf{A}(t,\mathbf{r}) &=&\int \frac{\mathbf{j(}t,\mathbf{r}_{1}\mathbf{)}%
+\partial _{t}\mathbf{D(}t,\mathbf{r}_{1}\mathbf{)}}{4\pi \left| \mathbf{r}-%
\mathbf{r}_{1}\right| }d^{3}r_{1},  \label{34}
\end{eqnarray}
respectively, and we have that $\mathbf{D}(t,\mathbf{r})=-\mathbf{\nabla }_{%
\mathbf{r}}\varphi (t,\mathbf{r})+\mathbf{\nabla }_{\mathbf{r}}\times
\mathbf{Z}(t,\mathbf{r})$ and $\mathbf{H}(t,\mathbf{r})=\mathbf{\nabla }_{%
\mathbf{r}}\times \mathbf{A}(t,\mathbf{r})+\mathbf{\nabla }_{\mathbf{r}}V(t,%
\mathbf{r})$. We obtain then, from Eq.(\ref{33}),

\begin{equation}
\mathbf{D}(t,\mathbf{r})=\frac{1}{4\pi }\int d^{3}r_{1}\frac{\rho (t,\mathbf{%
r}_{1})}{\left| \mathbf{r}-\mathbf{r}_{1}\right| ^{2}}\frac{\left( \mathbf{r}%
-\mathbf{r}_{1}\right) }{\left| \mathbf{r}-\mathbf{r}_{1}\right| }+\mathbf{%
\nabla }_{\mathbf{r}}\times \mathbf{Z}(t,\mathbf{r}).  \label{41a}
\end{equation}

For the special case of a point-like charge moving along the curve $\mathbf{r%
}_{0}(t)$ we put $\rho(t,\mathbf{r})=q\delta(\mathbf{r}-\mathbf{r}_{0}(t))$
and the above expression reduces to

\begin{equation}
\mathbf{D}(t,\mathbf{r})=\frac{q}{4\pi \left| \mathbf{r}-\mathbf{r}%
_{0}(t)\right| ^{2}}\frac{\mathbf{r}-\mathbf{r}_{0}(t)}{\left| \mathbf{r}-%
\mathbf{r}_{0}(t)\right| }+\mathbf{\nabla }_{\mathbf{r}}\times \mathbf{Z}(t,%
\mathbf{r}).  \label{41b}
\end{equation}

According to Eqs.(\ref{41a}) or (\ref{41b}) the field $\mathbf{D}(t,\mathbf{r%
})$ at time $t$ entails an instantaneous Coulomb field produced by a
continuous charge distribution $\rho $, or else by a point-like charge $q$.
Such a result would correspond to an instantaneous response of the field to
any change suffered by the charge distribution. That would be in
contradiction with the finite propagation-time needed by any signal.
Whatever the field $\mathbf{Z}(t,\mathbf{r})$ might be, it must contain a
similar instantaneous contribution that cancels the former one, if we want
the present approach to bear any physical interpretation. Such an issue has
been discussed and cleared, in the case of the \emph{complete} set of
Maxwell equations, by showing that both the longitudinal and the transverse
parts of the electric field contain instantaneous contributions, which turn
out to cancel each other \cite{donnely}. Note also that by taking $\mathbf{Z}
$ equal to zero in Eq.(\ref{40}) we have $\mathbf{\nabla }\times \mathbf{D=0}
$ in our case, which is not what happens when $\mathbf{D}$ has to satisfy
(together with $\mathbf{H}$) the complete system of Maxwell equations. In
any event, as we have already stressed, it is necessary to add some
additional information to the one derived from the continuity equation, in
order to fix the dynamics of the fields. We do this in the following form.
Instead of taking the potentials $\varphi $ and $\mathbf{A}$ as given by
Eqs.(\ref{33}) and (\ref{34}), we assume them as additional quantities, out
of which we define the fields $\mathbf{E}$ and $\mathbf{B}$ through

\begin{eqnarray}
\mathbf{E}(t,\mathbf{r}) &=&-\mathbf{\nabla }_{\mathbf{r}}\varphi (t,\mathbf{%
r})-\partial _{t}\mathbf{A}(t,\mathbf{r}),  \label{35} \\
\mathbf{B}(t,\mathbf{r}) &=&\mathbf{\nabla }_{\mathbf{r}}\times \mathbf{A}(t,%
\mathbf{r}).  \label{36}
\end{eqnarray}

These fields obey then the homogeneous Maxwell equations identically:

\begin{eqnarray}
\mathbf{\nabla \cdot B} &=&0  \label{37} \\
\mathbf{\nabla \times E}+\partial _{t}\mathbf{B} &=&\mathbf{0}.  \label{38}
\end{eqnarray}

Side by side to these two Maxwell equations we write the inhomogeneous ones:

\begin{align}
\mathbf{\nabla }\cdot \mathbf{D}& =\rho  \label{39} \\
\mathbf{\nabla }\times \mathbf{H}& =\mathbf{j}+\partial _{t}\mathbf{D.}
\label{40}
\end{align}

We stress once again that - up to this point - the homogeneous and the
inhomogeneous equations are independent from one another. We may connect
them through some \emph{constitutive equations}, like, e.g.,

\begin{align}
\mathbf{D}& =\varepsilon \mathbf{E,}  \label{41} \\
\mathbf{H}& =\mu ^{-1}\mathbf{B.}  \label{42}
\end{align}

These equations are usually assumed to describe a linear medium of
electrical permittivity $\varepsilon $ and magnetic permeability $\mu $. A
particular case of such a medium is vacuum, and the system of equations,
Eqs.(\ref{37}, \ref{38}, \ref{39}, \ref{40}), that arises out of a
connection like the one given by Eqs.(\ref{41}, \ref{42}) is what we know as
the complete system of Maxwell equations.

Without connecting $\left( \mathbf{D}\text{, }\mathbf{H}\right) $ with $%
\left( \mathbf{E}\text{, }\mathbf{B}\right) $ through some constitutive
equations, we have no closed system. The equations that we have written down
for $\left( \mathbf{D}\text{, }\mathbf{H}\right) $, that is Maxwell
equations with sources, can also be written down for a fluid, for example.
We can expect that any conclusion that can be derived in the realm of
electrodynamics from the equations $\mathbf{\nabla }\cdot \mathbf{D}=\rho $
and $\mathbf{\nabla }\times \mathbf{H}=\mathbf{j}+\partial _{t}\mathbf{D}$
\emph{without} coupling them to the source-free Maxwell equations, will have
a corresponding result in the realm of fluid dynamics. This assertion can be
illustrated by two examples: 1) A fluid having a point-like singularity in
its density $\rho $ (one can achieve this approximately, by using an
appropriate sink): one obtains in this case a velocity-field obeying a law
that is mathematically identical to Coulomb's law \cite{batchelor}. 2) A
fluid where a so-called vortex tube appears (tornadoes and whirl-pools are
associated phenomena), in which case - after approximating the vortex-tube
by a line singularity - one obtains a velocity-field through an expression
which is mathematically identical to the Biot-Savart law \cite{batchelor}.

\section{Tensorial formulation}

The derivation of the inhomogeneous Maxwell equations as a consequence of
charge conservation is nothing new \cite{parrot,hehl1,hehl2}. It follows as
a direct application of a theorem of de Rahm for differential forms \cite
{parrot,flanders,hehl1}. According to this theorem, given a four-vector $%
j^{\alpha }(x)$ for which a continuity equation holds, i.e., $\partial
_{\alpha }j^{\alpha }=0$, there exists an antisymmetric tensor $F^{\alpha
\beta }=-F^{\beta \alpha }$ fulfilling $\partial _{\alpha }F^{\alpha \beta
}=j^{\beta }$. As we said before, this last equation is nothing but the
tensorial form of the inhomogeneous Maxwell equations, Eqs.(\ref{39}) and (%
\ref{40}). Now, the tensor $F^{\alpha \beta }$ is not always derivable from
a vector $A^{\alpha }$. In order to be derivable from $A^{\alpha }$ in the
form $F^{\alpha \beta }=\partial ^{\alpha }A^{\beta }-\partial ^{\beta
}A^{\alpha }$, it must satisfy the equation $\partial ^{\alpha }F^{\beta
\gamma }+\partial ^{\beta }F^{\gamma \alpha }+\partial ^{\gamma }F^{\alpha
\beta }=0$. This is the tensorial form of the homogeneous Maxwell equations,
Eqs.(\ref{37},\ref{38}). In other words, given $j^{\alpha }$ and $A^{\alpha }
$, with $j^{\alpha }$ satisfying a continuity equation, we may introduce two
antisymmetric tensors, $F_{(1)}^{\alpha \beta }$ and $F_{(2)}^{\alpha \beta }
$. The first one can be determined so as to satisfy $\partial _{\alpha
}F_{(1)}^{\alpha \beta }=j^{\beta }$, according to de Rahm's theorem. The
second tensor, defined through $F_{(2)}^{\alpha \beta }\equiv \partial
^{\alpha }A^{\beta }-\partial ^{\beta }A^{\alpha }$, satisfies $\partial
^{\alpha }F_{(2)}^{\beta \gamma }+\partial ^{\beta }F_{(2)}^{\gamma \alpha
}+\partial ^{\gamma }F_{(2)}^{\alpha \beta }=0$ identically. In order that
these two equations do conform a closed system, i.e., the\emph{\ total }%
system of Maxwell equations, we need to connect $F_{(1)}^{\alpha \beta }$
with $F_{(2)}^{\alpha \beta }$ through some constitutive relation. In the
following we ellaborate on all this, but without employing the tools of
differential forms on manifolds, which - in spite of their usefulness -
cannot be said yet to be part of the lore of physics.

It is indeed not necessary to rest on de Rham's theorem and the theory of
differential forms on manifolds, in order to derive the foregoing
conclusions in tensorial form. One could start with the tensorial form of
Helmholtz theorem \cite{hauser,kobe1,kobe2} and go-ahead with a similar
reasoning as the one we have followed in the preceding sections. We shall
however proceed by explicitly constructing a tensor fulfilling our
requirements.

Let us thus start by assuming that we are given a vector field $j^\alpha $.
We want to show that there is an antisymmetric tensor $F^{\alpha \beta }$
fulfilling

\begin{equation}
\partial _{\alpha }F^{\alpha \beta }=j^{\beta }.  \label{t1}
\end{equation}

Note first that from Eq.(\ref{t1}) and the antisymmetry of $F^{\alpha \beta
} $ it follows that $j^{\beta }$ must satisfy the continuity equation:

\begin{equation}
\partial _\beta j^\beta =0.  \label{t5}
\end{equation}

We now demonstrate the existence of the tensor $F^{\alpha \beta }$ by
explicit construction. To this end, we consider the Green function $G\left(
x,x^{\prime }\right) $ satisfying

\begin{equation}
\partial _\mu \partial ^\mu G\left( x,x^{\prime }\right) =\delta \left(
x-x^{\prime }\right) .  \label{t4}
\end{equation}

Given $G\left( x,x^{\prime }\right) $ and $j^\alpha $ we introduce the
potential $A^\mu (x)$ as

\begin{equation}
A^\mu (x)=\int G\left( x,x^{\prime }\right) j^\mu (x^{\prime })d^4x^{\prime
},  \label{t2}
\end{equation}
and define

\begin{eqnarray}
F^{\mu \nu }\left( x\right) &\equiv &\partial ^{\mu }A^{\nu }(x)-\partial
^{\nu }A^{\mu }(x)  \label{t3} \\
&=&\int \left[ \partial ^{\mu }G\left( x,x^{\prime }\right) j^{\nu
}(x^{\prime })-\partial ^{\nu }G\left( x,x^{\prime }\right) j^{\mu
}(x^{\prime })\right] d^{4}x^{\prime }.
\end{eqnarray}

Let us now take the four-divergence of the above defined tensor $F^{\mu \nu
}\left( x\right) $:

\begin{equation}
\partial _\mu F^{\mu \nu }(x)=\int \left[ \partial _\mu \partial ^\mu
G\left( x,x^{\prime }\right) j^\nu (x^{\prime })-\partial _\mu \partial ^\nu
G\left( x,x^{\prime }\right) j^\mu (x^{\prime })\right] d^4x^{\prime }.
\label{t7}
\end{equation}

Because $G\left( x,x^{\prime }\right) $ satisfies Eq.(\ref{t4}), the first
integral in Eq.(\ref{t7}) is equal to $j^{\nu }(x)$. As for the second
integral, in order to show that it is zero we do as follows. Because $%
G\left( x,x^{\prime }\right) $ satisfies Eq.(\ref{t4}), it must be a
function of $\left( x-x^{\prime }\right) $, so that $\partial _{\mu }G\left(
x,x^{\prime }\right) =-\partial _{\mu }^{\prime }G\left( x,x^{\prime
}\right) $, where $\partial _{\mu }^{\prime }\equiv \partial /\partial
x^{\prime \mu }$. We use this property and integrate by parts the second
term in (\ref{t7}); at the same time we replace the first term by $j^{\nu
}(x)$:

\begin{eqnarray}
\partial _{\mu }F^{\mu \nu }(x) &=&j^{\nu }(x)+\partial ^{\nu }\int \left[
\partial _{\mu }^{\prime }\left( G\left( x,x^{\prime }\right) j^{\mu
}(x^{\prime })\right) -G\left( x,x^{\prime }\right) \partial _{\mu }^{\prime
}j^{\mu }(x^{\prime })\right] d^{4}x^{\prime } \\
&=&j^{\nu }(x)+\partial ^{\nu }\int \partial _{\mu }^{\prime }\left( G\left(
x,x^{\prime }\right) j^{\mu }(x^{\prime })\right) d^{4}x^{\prime }.
\label{t9}
\end{eqnarray}

We may now employ the generalized Gauss theorem in order to show that the
four-volume integral on the right-hand side of (\ref{t9}) vanishes. The
four-volume has as its boundary a three-dimensional hypersurface $S^{\prime }
$ whose differential element we denote by $dS_{\mu }^{\prime }$. Thus,
because $j^{\mu }$ vanishes at spatial infinity,

\begin{equation}
\int \partial _{\mu }^{\prime }\left( G\left( x,x^{\prime }\right) j^{\mu
}(x^{\prime })\right) d^{4}x^{\prime }=\oint G\left( x,x^{\prime }\right)
j^{\mu }(x^{\prime })dS_{\mu }^{\prime }=0,  \label{t10}
\end{equation}
when we let $S^{\prime }\rightarrow \infty $, and with this result Eq.(\ref
{t9}) reduces to (\ref{t1}).

Now, just as in the three-dimensional case, where the divergence of a field
did not determine the field uniquely (see Eq.(\ref{26})), by subjecting $%
F^{\mu \nu }$ to the sole condition of fulfilling Eq.(\ref{t1}) we do not
fix $F^{\mu \nu }$ completely. Indeed, the tensor $K^{\mu \nu }$, which is
defined below in terms of an \emph{arbitrary }four-vector $B_{\rho }$,
fulfills also Eq.(\ref{t1}):

\begin{eqnarray}
K^{\mu \nu } &=&F^{\mu \nu }-\frac{1}{2}\epsilon ^{\mu \nu \rho \sigma
}\left( \partial _{\rho }B_{\sigma }-\partial _{\sigma }B_{\rho }\right)
\label{tk} \\
&\equiv &F^{\mu \nu }-\frac{1}{2}\epsilon ^{\mu \nu \rho \sigma }H_{\rho
\sigma }\equiv F^{\mu \nu }-\widetilde{H}^{\mu \nu }.  \label{tk2}
\end{eqnarray}
Here, $\epsilon ^{\mu \nu \rho \sigma }$ is the totally antisymmetric
Levi-Civita tensor (in fact, a tensor density). The four-divergences of $%
K^{\mu \nu }$ and $F^{\mu \nu }$ are the same because, due to the
antisymmetry of $\epsilon ^{\mu \nu \rho \sigma }$ and the symmetry of
partial derivatives like $\partial _{\mu }\partial _{\rho }$, we have

\begin{equation}
\partial _{\mu }\widetilde{H}^{\mu \nu }=\frac{1}{2}\epsilon ^{\mu \nu \rho
\sigma }\left( \partial _{\mu }\partial _{\rho }B_{\sigma }-\partial _{\mu
}\partial _{\sigma }B_{\rho }\right) \equiv 0.
\end{equation}

Hence, we obtain Maxwell equation with sources:

\begin{equation}
\partial _{\alpha }K^{\alpha \beta }=j^{\beta },  \label{tm1}
\end{equation}
together with the identity $\partial _{\mu }{}F_{\nu \lambda }+\partial
_{\nu }{}F_{\lambda \mu }+\partial _{\lambda }{}F_{\mu \nu }\equiv 0$, which
follows from the definition of $F^{\mu \nu }$, as given in Eq.(\ref{t3}).
Introducing the dual $\widetilde{F}^{\mu \nu }=\epsilon ^{\mu \nu \alpha
\beta }F_{\alpha \beta }/2$ of the tensor $F_{\alpha \beta }$ we can write
the former identity in the form

\begin{equation}
\partial _{\mu }{}\widetilde{F}^{\mu \nu }=0.  \label{tm2}
\end{equation}

Eqs.(\ref{tm1}) and (\ref{tm2}) are Maxwell equations in tensorial form. As
we said before, they constitute a closed system as long as $K^{\alpha \beta
} $ and $F^{\mu \nu }$ become related to each other by some constitutive
equations. A general, linear algebraic, relationship between these tensors
has the form \cite{post}

\begin{equation}
K^{\alpha \beta }=\frac{1}{2}\chi ^{\alpha \beta \rho \sigma }F_{\rho \sigma
},  \label{tc1}
\end{equation}
where $\chi ^{\alpha \beta \rho \sigma }$ is called the constitutive tensor.
It has the following symmetry properties: $\chi ^{\alpha \beta \rho \sigma
}=-\chi ^{\beta \alpha \rho \sigma }=-\chi ^{\alpha \beta \sigma \rho }=\chi
^{\rho \sigma \alpha \beta }$. In three-dimensional notation the components
of $K^{\alpha \beta }$ are $\mathbf{D}$ and $\mathbf{H}$, whereas those of $%
F^{\mu \nu }$ are $\mathbf{E}$ and $\mathbf{B}$. For free-space, the only
nonzero components of $\chi ^{\alpha \beta \rho \sigma }$ have the values $%
\varepsilon _{0}$ and $1/\mu _{0}$, corresponding to the electrical
permittivity $\varepsilon _{0}$ and magnetic permeability $\mu _{0}$ of the
vacuum.

The properties of the medium can be specified through an equation like (\ref
{tc1}), as well as by introducing some other quantities that describe the
polarization and magnetization of the medium. In three-vector notation these
quantities are the vectors $\mathbf{P}$ and $\mathbf{M}$, respectively.
Their relation to $(\mathbf{D},\mathbf{H})$ and $(\mathbf{E},\mathbf{B})$ is
given, in the simplest case, by

\begin{eqnarray}
\mathbf{D} &=&\varepsilon _{0}\mathbf{E}+\mathbf{P},  \label{d} \\
\mathbf{H} &=&\frac{1}{\mu _{0}}\mathbf{B}-\mathbf{M}.  \label{h}
\end{eqnarray}

In tensor notation, $\mathbf{P}$ and $\mathbf{M}$ are subsumed into an
antisymmetric tensor:

\begin{equation}
M^{\alpha \beta }=\left(
\begin{tabular}{llll}
$0$ & $P_{1}$ & $P_{2}$ & $P_{3}$ \\
$-P_{1}$ & $0$ & $-M_{3}$ & $M_{2}$ \\
$-P_{2}$ & $M_{3}$ & $0$ & $-M_{1}$ \\
$-P_{3}$ & $-M_{2}$ & $M_{1}$ & $0$%
\end{tabular}
\right) .
\end{equation}
This choice corresponds to the assignment $E_{i}=F_{0i}$ for the electric
field, and $B_{i}=-\epsilon _{ijk}F_{jk}/2$ for the magnetic field (Latin
indices run from $1$ to $3$). By relating $M^{\alpha \beta }$ to $K^{\alpha
\beta }$ through $K^{\alpha \beta }=F^{\alpha \beta }-M^{\alpha \beta }$ we
can rewrite the inhomogeneous Maxwell equation (\ref{tm1}) as

\begin{equation}
\partial _{\alpha }F^{\alpha \beta }=j^{\beta }+\partial _{\alpha }M^{\alpha
\beta }.
\end{equation}
Written in this form, the inhomogeneous Maxwell equation makes the
magnetization-polarization tensor $M^{\alpha \beta }$ appear as a source of
the electromagnetic field $F^{\alpha \beta }\equiv \partial ^{\mu }A^{\nu
}(x)-\partial ^{\nu }A^{\mu }(x)$. The constitutive equations are given in
this case as a connection between $M^{\alpha \beta }$ and $F^{\alpha \beta }
$ \cite{pellegrini}. At any rate, one has to make some hypothesis concerning
the electromagnetic properties of the medium - be it vacuum or any other
kind of medium - in order to obtain the closed system of Maxwell equations.
The simplest assumption is to attribute to the medium the property of
reacting locally and instantaneously to the presence of a field. It could
be, however, that such an assumption describes reality as a first
approximation only.

Finally, we want to stress the central role played by the Green function $%
G(x,x^{\prime })$. We assumed this function to satisfy Eq.(\ref{t4}), an
equation entailing the velocity of light. One possible solution of (\ref{t4}%
) is given by the retarded Green function

\begin{equation}
G(x,x^{\prime })=\frac{1}{4\pi \left| \mathbf{r}-\mathbf{r}^{\prime }\right|
}\delta (t-t^{\prime }-\left| \mathbf{r}-\mathbf{r}^{\prime }\right| /c).
\label{t11}
\end{equation}
This is the solution of Eq.(\ref{t4}) to which we ascribe physical meaning.
By using it in Eq.(\ref{t2}) we are actually prescribing how the source $%
j^{\mu }(x^{\prime })$ at a space-time point $x^{\prime }$ gives rise to an
electromagnetic field $A^{\mu }(x)$ at a distant point $x$; a field that
virtually acts upon a second charge or current density that is located at
such a distant point. There is therefore a fundamental piece of information
concerning the electromagnetic properties of space-time that is already
contained in the Green function, be it given through the special form of $%
G(x,x^{\prime })$, as in Eq.(\ref{t11}), or through the equation it has to
satisfy, e.g., Eq.(\ref{t4}).

\section{Summary and Conclusions}

We showed that starting from charge conservation one can arrive at equations
which are mathematically identical to Maxwell equations with sources. These
equations are therefore tightly linked to a general statement telling us
that something is conserved. Consider anything - charge, matter, or whatever
- that is contained inside an arbitrary volume. Consider also that this
thing is in a quantity that changes with time. If the change is exclusively
due to a flow through the volume's boundary, a continuity equation holds
true. Then, as a consequence of it, a pair of Maxwell-like equations must be
fulfilled by some auxiliary fields, which take the role ascribed to $\mathbf{%
D}$ and $\mathbf{H}$ in Maxwell equations.

That Maxwell equations with sources follow from charge conservation is a
mathematical fact that has been known since a couple of years \cite
{parrot,hehl1}, although it is usually not mentioned in standard textbooks
of electromagnetism. Maxwell equations with sources involve the fields $%
\mathbf{D}$ and $\mathbf{H}$, whereas the source-free equations involve the
fields $\mathbf{E}$ and $\mathbf{B}$. It is through some constitutive
equations connecting $\left( \mathbf{D}\text{, }\mathbf{H}\right) $ with $%
\left( \mathbf{E}\text{, }\mathbf{B}\right) $ that we obtain a closed
system, i.e., the complete system of Maxwell equations. The constitutive
equations express, in some way or another, the underlying properties of the
medium where the fields act or are produced. From this perspective, the
Maxwell equations entail besides charge conservation some properties of the
medium, yet to be unraveled. These properties are effectively described, in
the simplest case, through the permittivity $\varepsilon $ and the
permeability $\mu $ of the medium. The first one refers to electrical, the
second one to magnetic, properties of the medium, be it vacuum or any other
one. It is just when the equations for $\left( \mathbf{D}\text{, }\mathbf{H}%
\right) $ together with those for $\left( \mathbf{E}\text{, }\mathbf{B}%
\right) $ do conform a closed system, that we can derive a wave equation for
these fields. The velocity of wave propagation is then given by $%
c=(\varepsilon \mu )^{-1/2}$, the velocity of light. This must be in
accordance with the assumptions we make when choosing a physically
meaningful Green function. It is remarkable that the velocity of light can
be decomposed in terms of a product of two independent parameters. However,
the development of physics has led us to see $c$ as a fundamental constant
of Nature, instead of $\varepsilon $ and $\mu $. Nevertheless, currently
discussed and open questions related to accelerated observers, Unruh
radiation, self-force on a charge, magnetic monopoles and the like, might
well require an approach where the role of $c$ recedes in favor of
quantities like $\varepsilon $ and $\mu $. Maxwell equations, when written
in the - by now - most commonly used Gaussian units, do not include but the
single constant $c$, hiding so $\varepsilon $ and $\mu $ from our view.
These last two constants might well be key pieces that remain buried under
the beauty of a unified theory of electromagnetic phenomena, which is the
version of electrodynamics that we know and use today. A version that should
not be regarded as a closed chapter in the book of classical physics.

\section{Acknowledgments}

The author is very much indebted to Professor F. W. Hehl for his
comments concerning a first version of the present article, as
well as for drawing his attention to the rich literature on a
series of topics related to the present work.

\end{document}